\documentclass[doublecol]{epl2} 

\usepackage{hyperref}
\hypersetup{colorlinks=true, urlcolor=blue}

\usepackage{epsfig}
\usepackage{newlfont}
\usepackage{amssymb}
\usepackage{amsfonts}
\usepackage{amsmath}
\usepackage{bm}

\def\lsim{\mathrel{\rlap{\lower4pt\hbox{\hskip1pt$\sim$}}
    \raise1pt\hbox{$<$}}}                
\def\gsim{\mathrel{\rlap{\lower4pt\hbox{\hskip1pt$\sim$}}
    \raise1pt\hbox{$>$}}}                

\title{Quantum discord surge heralds entanglement revival in an infinite spin chain}
\shorttitle{Quantum discord surge heralds entanglement revival} 

\author{Himadri Shekhar Dhar\inst{1} \and Rupamanjari Ghosh\inst{1} \and Aditi Sen(De)\inst{2} \and Ujjwal Sen\inst{2}}
\shortauthor{H.S. Dhar \etal}

\institute{
  \inst{1} School of Physical Sciences, Jawaharlal Nehru University, New Delhi 110067, India\\
  \inst{2} Harish-Chandra Research Institute, Chhatnag Road, Jhunsi, Allahabad 211019, India
}


\abstract{A measure of quantum correlation defined from an information-theoretic perspective, namely, quantum discord, 
is applied to study the time-evolved nonequilibrium state of the infinite anisotropic quantum XY spin chain in a transverse time-dependent field. In particular, we probe whether the collapse and revival of nearest-neighbor entanglement of the state seen with a varying initial applied field strength, at a fixed evolution time, may be predicted from the behavior of the quantum correlation measure. For this quantum many-body system, realizable with currently available technology, 
we find that the revival of entanglement of the evolved state happens if there is an increase in quantum discord in the vicinity of entanglement collapse.}

\begin{document}

\maketitle
\section{Introduction}

There have been extensive studies in many-body and quantum information theory (QIT), attempting to quantify the key aspects of quantum correlations present between the parts of a system. The foremost among them is on the entanglement-separability distinction and its subsequent quantification by using measures such as entanglement of formation, distillable entanglement, and relative entropy of entanglement (see \cite{ent} for a review). Entanglement has been extensively used for indicating quantum criticality in phase transitions \cite{fazio} and in numerous QIT applications such as quantum teleportation \cite{tele}, quantum dense coding \cite{dense-coding}, and quantum key distribution \cite{qkd}. However, there is no quantitative agreement between the various measures of entanglement \cite{ent}. Even qualitative differences appear: Distillable entanglement vanishes for a certain class of entangled states, called bound entangled states \cite{bound-ent}, although other entanglement measures produce nonzero values. 

Moreover, there exist quantum correlations which appear even when entanglement is absent. Several phenomena have been discovered which produce nonclassical results with no shared entanglement identified in the system. These correlations feature in important aspects of QIT, such as ``quantum non-locality without entanglement" \cite{nlwe} (cf. \cite{disting}) and ``quantum data hiding in separable states" \cite{Vincenzo}. A natural question then arises: What form of quantum correlations is responsible for such nonclassical (nonlocal) behavior even in the absence of entanglement? 

Early attempts to explore the concept of nonclassicality and correlations from a perspective that is different from the entanglement-separability paradigm include those defined in the language of quantum optics \cite{quantumopt}, and the literature on Bell inequalities \cite{Bell}. Recently, information-theoretic and thermodynamic concepts have been used to define quantum correlations independent of the entanglement-separability criterion -- the measures of $\textit{quantum discord}$ (QD) \cite{zur,ved} and $\textit{quantum work-deficit}$ (QWD) \cite{wd} quantify quantum correlations by attempting to quantize expressions for correlations existing in classical information theory. The mapping from a classical to a quantum system often introduces a lack of concord, mainly due to the non-commutativity of quantum operators. QD quantifies quantum correlations by using the difference in the quantum expressions corresponding to two equivalent definitions of classical mutual information \cite{Cover-Thomas}. QWD, on the other hand, does it by using the difference in the amount of negentropy (``work'') extractable by global and local heat engines \cite{thermo-book, Cerf}. The application of such measures of quantum correlations in many-body systems may reveal new phenomena which cannot be detected by entanglement. QD has already been applied for studying static properties, such as properties of the ground state and the equilibrium states, of quantum spin systems \cite{discord-spin}. There is currently no concrete operational relation between entanglement and other measures of quantum correlations such as QD or QWD.

In order to probe this interrelation, we focus here on a known, exactly solvable model, that of an infinite anisotropic XY (quantum) spin chain in a transverse field. The nearest-neighbor entanglement in the time-evolved state of the chain, at a given fixed time, exhibits critical behavior -- a dynamical phase transition (DPT), controlled by the initial value $a$ of the transverse field \cite{5a}. We study the dynamics of quantum correlations, as quantified by QD between two neighboring spins, and show that, for this particular system, the observed collapse of entanglement and higher-field revival (with changing initial value $a$ of the transverse field) can be predicted by the behavior of QD -- entanglement revives after collapse, if QD increases in the vicinity of the region where entanglement collapse occurs \footnote{The collapse and revival of entanglement occurs with the changing initial value $a$ of the transverse field for a fixed evolution time.}. Hence, an increasing QD at the point of entanglement collapse (i.e., at the corresponding time and field strength) can be used as an indicator for the entanglement revival. We further show that this behavior is potentially generic in that it does not depend on the specific measure of quantum correlation used -- QWD also shows a similar predictive capacity, although further work is needed to extract a straightforward criterion \cite{inprep}.

\section{Measures of quantum correlation}
For completeness, we first define the correlation measures.
\subsection{Quantum discord}

The total correlation in any bipartite quantum system, $\rho_{AB}$ shared between two parties \(A\) and \(B\), can be measured by using quantum mutual information \cite{qmi} (see also \cite{Cerf, GROIS}):
\begin{equation}
\label{qmi1}
I(\rho_{AB})= S(\rho_A)+ S(\rho_B)- S(\rho_{AB}) ,
\end{equation}
where $S(\rho)= - \mbox{tr} (\rho \log_2~ \rho)$ is the von Neumann entropy of the quantum state \(\rho\), and \(\rho_A\) and \(\rho_B\) are the local density matrices of \(\rho_{AB}\). 

Classical mutual information can be similarly defined for a joint probability distribution \(\{p_{ij}\}\) as
\(I(\{p_{ij}\}) = H(\{p_{i.}\}) + H(\{p_{.j}\}) - H(\{p_{ij}\})\),
where \(H(\{q_j\}) = -\sum_j q_j \log_2 q_j\) is the Shannon entropy of the probability distribution \(\{q_j\}\), and \(\{p_{i.}\}\), \(\{p_{.j}\}\) are the marginals of \(\{p_{ij}\}\). 
There is an equivalent classical expression for mutual information using the concept of conditional entropy:
\(H(\{p_{ij}\})= H(\{p_{.j}\})+ H(\{p_{i|j}\}) = H(\{p_{i.}\})+ H(\{p_{j|i}\})\),
where \(\{p_{i|j}\}\) and \(\{p_{j|i}\}\) are the conditional probability distributions. 

For our bipartite quantum system, measuring on the subsystem \(B\) using the set of projectors \(\{B_i\}\) [where \(B_iB_j = \delta_{ij} B_i\), \(\sum_i B_i = \mathbb{I}_B\), with \(\mathbb{I}_B\) being the identity operator on the Hilbert space on which \(\rho_B\) is defined], when the two-particle system is in the quantum state \(\rho_{AB}\), produces the post-measurement states \(\rho^{i}_{AB} = \frac{1}{p_i} \mathbb{I}_A \otimes B_i \rho \mathbb{I}_A \otimes B_i\), where \(\mathbb{I}_A\) is the identity operator on subsystem \(A\), and \(p_i = \mbox{tr}_{AB}(\mathbb{I}_A \otimes B_i \rho \mathbb{I}_A \otimes B_i)\). The conditional quantum states that are produced at \(A\), due to the measurement at \(B\), are 
\(\rho_{A|i} = \frac{1}{p_i} \mbox{tr}_B(\mathbb{I}_A \otimes B_i \rho \mathbb{I}_A \otimes B_i)\), with probability \(p_i\). 
The quantum conditional entropy can then be defined as 
\( S(\rho_{A|B}) = \min_{\{B_i\}} \sum_i p_i S(\rho_{A|i})\),
and similarly,
\(S(\rho_{B|A})\). 
One can quantize the classical expression 
\begin{equation}
 I(\{p_{ij}\}) = H(\{p_{i.}\}) - H(\{p_{i|j}\})
\end{equation}
for mutual information to obtain the following:
\begin{equation}
 J(\rho_{AB}) = S(\rho_A) - S(\rho_{A|B}).
\end{equation}
QD is then defined as
\begin{equation}
Q(\rho_{AB})= I(\rho_{AB}) - J(\rho_{AB}).
\end{equation}
QD is positive for all quantum states. 

\subsection{Quantum work-deficit}
The concept of QWD is based on the fact that information can be treated as a thermodynamic resource \cite{info-resource}. Given a quantum state \(\rho\), 
one defines the allowed class of global quantum operations, called ``closed operations'' (CO), as arbitrary sequences of 
the following operations: (G1) unitary operations and (G2) dephasing \(\rho\) by using a set of projectors \(\{P_i\}\), i.e., \(\rho \rightarrow \sum_i P_i \rho P_i\),  
where \(P_iP_j = \delta_{ij} P_i\), \(\sum_i P_i = \mathbb{I}\), with 
\(\mathbb{I}\) being the identity operator on the Hilbert space \({\cal H}\) on which \(\rho\) is defined. Under this class of operations, it can be shown (see \cite{wd, horohuge}) that the number of pure qubits that can be extracted from \(\rho\) is 
\(I_G (\rho)= N - S(\rho)\), where \(N = \log_2 \dim {\cal H}\).

Correspondingly, the allowed class of local operations is called ``closed local operations and classical communication'' (CLOCC), and is defined as arbitrary compositions of the following operations: (L1) local unitary operations, (L2a) local dephasing and (L2b) sending a
completely dephased subsystem from one party to another over a noiseless quantum channel. Let us consider a bipartite quantum state \(\rho =\rho_{AB}\). The number of qubits that can be extracted from a bipartite quantum state \(\rho_{AB}\) under CLOCC is 
\begin{equation}
I_L(\rho_{AB}) = N - \inf_{\Lambda \in CLOCC} [S(\rho{'}_A) + S(\rho{'}_B)],
\end{equation}
where \(\rho{'}_A = \mbox{tr}_B (\Lambda (\rho_{AB}))\), \(\rho{'}_B = \mbox{tr}_A (\Lambda (\rho_{AB})\), and now \(N = \log_2 \dim {\cal H}_{AB}\), with \({\cal H}_{AB}\) being the Hilbert space on which \(\rho_{AB}\) is defined.
QWD is then defined as 
\begin{equation}
 \Delta(\rho_{AB}) = I_G(\rho_{AB}) - I_L(\rho_{AB}).
\end{equation}

\subsection{Entanglement: Logarithmic negativity}
There is a plethora of entanglement measures that is known in the community, and each has its own operational relevance and degree of mathematical tractability. We will use the logarithmic negativity (LN), which is a very useful computable measure of entanglement \cite{9a}.

The definition of LN is based on the fact that the negativity of the partial transpose of a bipartite quantum state is a sufficient condition for the state to be entangled -- the Peres-Horodecki separability criterion \cite{10a}. Moreover, for two-qubit systems, which will be our domain of study, the condition is necessary and sufficient \cite{Horo-1}.  

LN is evaluated by using a quantity called ``negativity'', defined for the state \(\rho_{AB}\) as 
\begin{equation}
\cal{N}(\rho_{AB}) = \frac{\left\|\rho_{AB}^{T_A}\right\|_1- 1}{2},
\end{equation}
where $\left\|\rho_{AB}^{T_A}\right\|_1$ is the trace norm of the partial transpose \(\rho_{AB}^{T_A}\) of \(\rho_{AB}\). 
From the Peres-Horodecki separability criterion, the partial transpose $\rho_{AB}^{T_A}$ should be positive for all separable states. Hence $\cal{N}(\rho_{AB})$ is zero for separable states. 
LN of \(\rho_{AB}\) is defined as 
\begin{eqnarray}
E_{\cal{N}}(\rho_{AB})& = & \log_2 \left\|\rho_{AB}^{T_A}\right\|_1 \nonumber \\
&\equiv & \log_2 [ 2 \cal{N}(\rho_{AB}) +1 ].
\end{eqnarray}

\section{The system: infinite quantum XY spin chain in a transverse field}
The infinite anisotropic XY spin chain in a transverse field is governed by the Hamiltonian,
\begin{equation}
\textsl{H}= J\sum_i [(1+\gamma)S^x_iS^x_{i+1} + (1-\gamma)S^y_iS^y_{i+1}] - h(t)\sum_i S^z_i ,
\end{equation}
where the anisotropy \(\gamma\) is nonzero, and \(J\) measures the interaction strength. $S^{j} = \frac{1}{2}\sigma^j$ ($\textit{j=x,y,z}$) are one-half of 
the Pauli spin matrices at the corresponding site. Note that \(\gamma = 0\) corresponds to the XX model while \(\gamma = 1\) is for the Ising model. Here we consider the models for \(\gamma > 0\), so that the interaction and the field parts of the Hamiltonian do not commute, whereby the external field can have nontrivial effects on the evolution. The transverse field is applied in the form of an initial disturbance:
\begin{equation}
h(t)= \begin{cases}
              a,~~~t~=~0 \\
              0,~~~t~\geq~0,  
      \end{cases}
\end{equation}  
where $a \neq 0$.

The above Hamiltonian can be realized in a system of cold atoms confined in an optical lattice. The two-component Bose-Bose and Fermi-Fermi mixtures, in the strong coupling limit with suitable tuning of scattering length and additional tunneling in the system can be described by the above Hamiltonian \cite{Sachdev,fazio}. The dynamics of the system can be simulated by controlling the system parameters and the applied transverse field \cite{ripoll}.

Suppose that the system starts off from the initial state which is a (canonical) equilibrium state at temperature T. We are interested in the nearest-neighbor (two-site) density matrix of the evolved state at time \(t\), that started off from the equilibrium state. In general, a two-qubit density matrix is of the form
\begin{eqnarray}
& \frac{1}{4} \Big(I \otimes I + \sum_{j=x,y,z} M^j (t) ( \sigma^j \otimes I + I \otimes \sigma^j) \nonumber \\
& + \sum_{j,k=x,y,z} T^{jk}(t) \sigma^j \otimes \sigma^k \Big), \nonumber
\end{eqnarray}
where \(T^{jk}(t)\) are the two-site correlation functions, and \(M^j(t)\) are the magnetizations. Using properties of the XY Hamiltonian, some simplifications can be made, and the final form of the two-site density matrix is \cite{6a,7a,8a}
\begin{eqnarray}
\rho^{\textit{12}}_{\beta}(t)&=& \frac{1}{4} \Big( I \otimes I + M^z (t)( \sigma^z \otimes I + I \otimes \sigma^z) \nonumber \\ 
&& + \quad  T^{xy}(t)(\sigma^x \otimes \sigma^y + \sigma^y\otimes \sigma^x) \nonumber \\
&& +\sum_{j=x,y,z} T^{jj}_\beta(t) \sigma^j \otimes \sigma^j \Big).
\end{eqnarray}
Diagonalizing the Hamiltonian via Jordan-Wigner and Fourier transformations, the correlations and the transverse magnetization in (11), for an initial temperature T = 0, are found to be \cite{6a,7a,8a}: 
\(T^{xy}(t)= T^{yx}(t)=S(t)\), \(T^{xx}(t)=G(-1,t)\),
\(T^{yy}(t)=G(1,t)\),
and 
\(T^{zz}(t)= [M^z(t)]^2-G(1,t)G(-1,t) + [S(t)]^2 \),
where $G(R,t)$ (for $R~=\pm~1$), $S(t)$ are given by
\begin{eqnarray}
G(R,t) &=&  \frac{\gamma}{\pi}\int^\pi_0d\phi~\sin(\phi R)\sin \phi \frac{1} {\Lambda(\tilde{a})\Lambda^2(0)}\nonumber\\\nonumber\\
& &\times\left\{\gamma^2 \sin^2\phi~+~(\cos\phi-\tilde{a})\cos\phi \right.  \nonumber\\
& &+ \left. \tilde{a}\cos\phi\cos[2\Lambda(0)\tilde{t}]\right\} \nonumber\\\nonumber\\
& &-\frac{1}{\pi}\int^\pi_0d\phi~\cos\phi \frac {1} {\Lambda(\tilde{a}) \Lambda^2(0)}\nonumber\\\nonumber\\
& & \times \left( \left\{\gamma^2\sin^2\phi+\left(\cos\phi-\tilde{a} \right)\cos\phi\right\} \right. \cos\phi\nonumber\\
& & \left. -~\tilde{a}\gamma^2\sin^2\phi\cos[2\Lambda(0)\tilde{t}] \right),
\nonumber
\end{eqnarray}
\begin{equation}
S(t) = - \frac{\gamma \tilde{a}}{\pi} \int^\pi_0d\phi \sin^2 \phi \frac{\sin[2\tilde{t}\Lambda(0)]}{\Lambda(\tilde{a}) \Lambda(0)} ,
\nonumber
\end{equation}
and 
\begin{eqnarray}
M^z(t) &=& \frac{1}{\pi} \int_0^\pi d\phi \frac{1}{\Lambda(\tilde{a}) \Lambda^2(0)} \nonumber \\
     &\times& \{\cos[2 \Lambda(0)\tilde{t}] \gamma^2 \tilde{a}\sin^2\phi] \nonumber \\
  &-& \cos\phi [(\cos\phi - \tilde{a})\cos\phi + \gamma^2 \sin^2 \phi]\}. \nonumber
\end{eqnarray}
Here 
\(\Lambda(x)= \left\{\gamma^2\sin^2\phi~+~[x-\cos\phi]^2\right\}^{\frac{1}{2}}\),
and 
\( \tilde{a} = a/J, \quad \tilde{t} = Jt/\hbar\).
We will use \(\tilde{a}\) and \(\tilde{t}\) as the (dimensionless) initial field and time parameters, respectively.

\section{Methodology}
We observe the behavior of the system at a time \(\tilde{t}\) after it starts from an initial canonical equilibrium state at zero temperature. In particular, we wish to find the measurement strategy for obtaining the optimal QD for the nearest-neighbor (two-qubit) density matrix of the evolved state at time \(\tilde{t}\). Since we focus on projection-valued measurements, and since the local systems are qubits, the measurement will necessarily involve projecting onto an orthonormal (two-element) basis of a two-dimensional complex Hilbert space. Let that basis be given by 
\begin{eqnarray}
\label{eq-horibol}
\left|i_1\right\rangle &=& \cos \frac{\theta}{2} \left|0\right\rangle + e^{i\phi} \sin \frac{\theta}{2} \left|1\right\rangle , \nonumber \\
\left|i_2\right\rangle &=&  e^{-i\phi} \sin \frac{\theta}{2} \left|0\right\rangle + \cos \frac{\theta}{2} \left|1\right\rangle ,
\end{eqnarray}
where \(\{|0\rangle, |1\rangle\}\) form the computational qubit basis.

In the case of QWD, the general definition involves two-way communication of dephased states, and is as yet not computable for arbitrary states. We consider the restricted case where only one-way communication is allowed. Again, as the local subsystems are qubits, the most general measurement basis for the dephasing will be of the form in Eq.\,(\ref{eq-horibol}).

In our analysis, we consider the state of the system as we sweep over the applied initial field strength $\textit{$\tilde{a}$}$, at a fixed value of the anisotropy parameter $\gamma$ for a fixed time of evolution of the zero-temperature equilibrium state. The motivation for considering such a state is the fact that the anisotropic transverse XY model, at zero temperature, undergoes a quantum phase transition (QPT) at $\textit{$\tilde{a}$}$ = 1, as we sweep over the magnetic field at a fixed $\gamma$. An initial zero-temperature equilibrium state ensures that thermal fluctuations are absent. 

\section{Entanglement versus Quantum Discord}
\begin{figure}[htb]
\begin{center}
\epsfig{figure=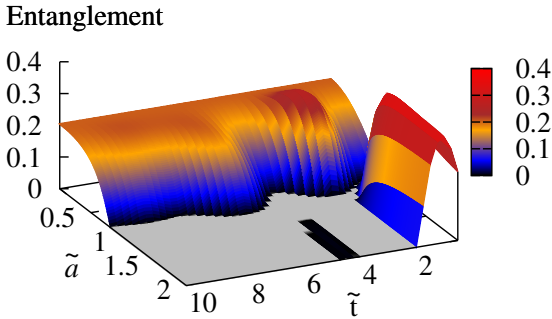, height=.3\textheight, width=.23\textwidth,angle=-90}
\end{center}
\caption{(Color online) Behavior of entanglement (as quantified by logarithmic negativity, measured in ebits) with respect to 
the time of evolution \(\tilde{t}\) (dimensionless) and the initial field strength \(\tilde{a}\) (dimensionless). The collapse of entanglement to zero are
 distinctly shown in the regions around \(\tilde{t}=1\) and \(\tilde{t}=4\). We have taken \(\gamma = \frac{1}{2}\).}
\label{fig:1}
\end{figure}

%
We consider the initial states to be equilibruim states at zero temperature. The observed DPTs occur for a wide range of the asymmetry parameter $\gamma$. In Fig.\,\ref{fig:1}, we plot LN of the nearest-neighbor state of the evolved state, as a function of time 
$\tilde{t}$ and the initial field strength $\tilde{a}$. An important observation from the figure is that for times until a little before \(\tilde{t} = 2\), and again at around \(\tilde{t} = 4\), the entanglement collapses to zero at a certain field value, but revives to give nonzero values at higher fields in the dynamic evolution. At other times, the entanglement recedes to zero, and does \emph{not} become nonzero at higher field values. This remarkable behavior of collapse and higher-field revival of entanglement in this case takes place with respect to the varying initial field applied to the system \footnote{A temporal collapse and revival of entanglement can also be observed in Fig.\,1. The entanglement collapses and revives with varying time for certain values of the initial field strength ($\tilde{a} > 1.2$). Our primary interests, however, lie at understanding the correlations that arise close to the zero-temperature QPT at a fixed time of evolution of the quantum state. Note that the QPT at zero temperature also happens at a fixed time, which is \(\tilde{t}=0\), as we sweep over the \(\tilde{a}\) axis. The DPT, observed in Ref.\,\cite{5a}, considers the status of the zero-time transition at nonzero times, if we still sweep over the \(\tilde{a}\) axis.}. For definiteness, the figure is plotted with the anisotropy \(\gamma = \frac{1}{2}\). However, all the results hold irrespective of the value of \(\gamma\) chosen in the range \((0,1]\). 

\begin{figure}[htb]
\begin{center}
\epsfig{figure=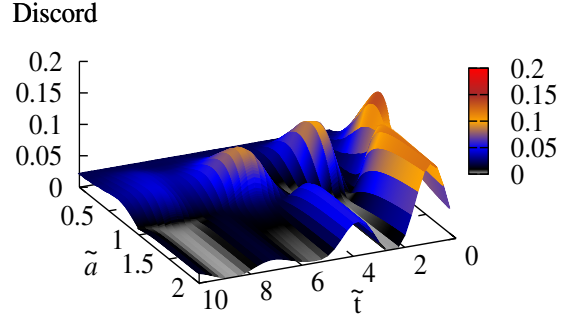, height=.3\textheight, width=0.23\textwidth,angle=-90}
\end{center}
\caption{(Color online) Behavior of quantum discord (measured in bits) as a function of time \(\tilde{t}\) (dimensionless) and initial field strength \(\tilde{a}\) (dimensionless). \(\gamma = \frac{1}{2}\), as before. QD shows a different behavior around times \(\tilde{t}=1\) and \(\tilde{t}=4\) for which entanglement reappearance takes place. This is made clearer in Fig.\,\ref{fig:3}.}
\label{fig:2}
\end{figure}

How does QD behave in this interesting range of initial field strength \(\tilde{a}\) and time \(\tilde{t}\)? The first observation in our computed behavior of QD shown in Fig.\,\ref{fig:2}
is that QD is nonzero at points where entanglement is zero. We further observe that the behavior of QD at points where entanglement revival occurs is markedly different. 

The system thus supports another form of quantum correlation, even when entanglement goes to zero. This leads us to pose the query: 
\emph{Are there more general forms of quantum correlations in the system, whose presence at a point of entanglement collapse, can predict the revival of entanglement for certain times of evolution, and whose absence anticipates a non-revival?} 
We give an affirmative answer to this question and formulate a relation that ascertains the observed behavior 
for the considered system. 
\emph{Consider a time-evolved nonequilibrium bipartite quantum state \(\rho_{AB}^a(\tilde{t})\), for an anisotropic XY spin chain in a transverse field, obtained by time-evolution for a duration \(\tilde{t}\) 
and varying with a system parameter \(\tilde{a}\). If for a fixed time \(\tilde{t}\), the entanglement \(E\) vanishes at \(\tilde{a}=\tilde{a}_c\), then}
\begin{equation}
\label{eita-asol-byapar}
\tilde{a}\frac{\partial Q(\rho_{AB})}{\partial \tilde{a}}\Big|_{\sim \left.\right. \tilde{a}_c} > 0 \implies E(\rho_{AB}) > 0 \mbox{ for some } |\tilde{a}| > |\tilde{a}_c|.
\end{equation}
Here \(Q\) stands for quantum discord 
\footnote{For the specific model, the violation of the inequality in (\ref{eita-asol-byapar}), usually results in no revival of entanglement. There may, however, exist few intermediary states  where entanglement revival occurs in absence of distinct positivity and the collapse-revival behavior of entanglement remains inconclusive. For positive \(a\) such resurgence of entanglement without positivity of (\ref{eita-asol-byapar}) around \(a=a_c\), happens in the vicinity of \(\tilde{a} = 1\), and so it is plausible that such exceptional cases are related to the zero-temperature QPT in this model \cite{Sachdev}.}
and the parameter \(\tilde{a}\) is the initial transverse field. From (\ref{eita-asol-byapar}) it is evident that the QD decreases if one considers DPT along the negative \(\tilde{a}\) axis. This is due to the reflection symmetry and surge (dip) in QD is along the positive (negative) direction of the considered physical parameter. Hence we consider just the positive parameter axis and the surge of QD.

\begin{figure}[htb]
\begin{center}
\epsfig{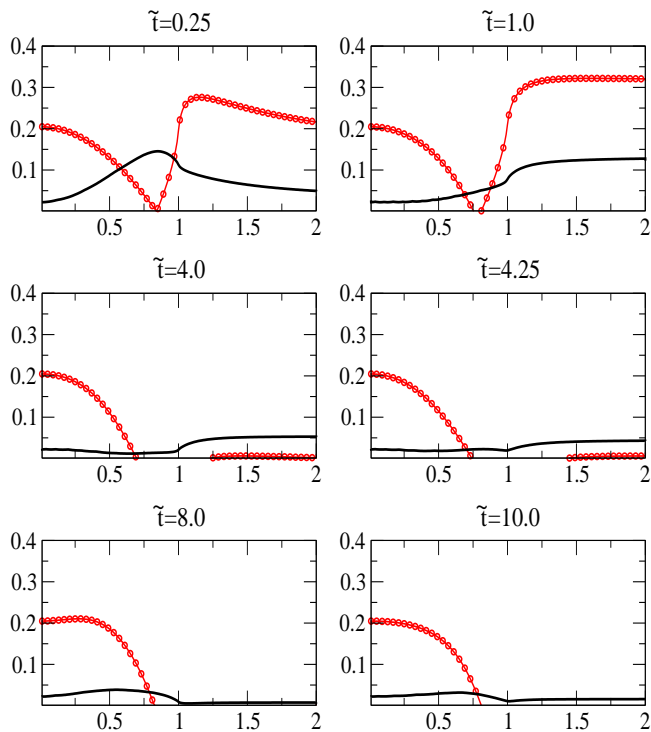}
\end{center}
\caption{(Color online) Quantum discord (continuous, black) and entanglement (logarithmic negativity) (beaded, red) for fixed times \(\tilde{t}\) as functions of the initial field strength \(\tilde{a}\). Entanglement collapses to zero and revives at higher-fields for times before \(\tilde{t} \approx 2\) and at around \(\tilde{t} = 4\). For these and only these times, QD increases around the collapse of LN. LN is measured in ebits, QD in bits, while the horizontal axes denoting the initial field strength \(\tilde{a}\) are dimensionless. \(\gamma = \frac{1}{2}\), as before.}
\label{fig:3}
\end{figure}

\begin{figure}[htb]
\begin{center}
\epsfig{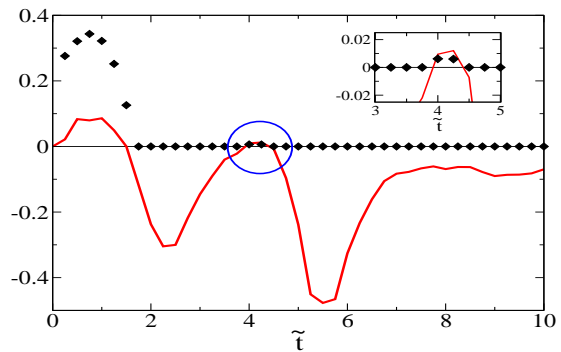}
\end{center}
\caption{(Color online) Increase of quantum discord at entanglement collapse indicates entanglement revival. The continuous red curve is of the (first) partial derivative of QD (measured in bits) with respect to the initial field strength at \(\tilde{a} = \tilde{a}_c\) versus time \(\tilde{t}\) (dimensionless). The black diamonds denote the maximum values attained by LN after its collapse (measured in ebits) versus \(\tilde{t}\). The continuous curve crosses over zero for \(\tilde{t} \lsim  2\), and again at \(\tilde{t} \approx 4\), which are exactly the times for which entanglement revival happens, i.e., where the curve of diamonds is nonzero. The inset magnifies the crossing over zero around \(\tilde{t} \approx 4\). Thus the maximum entanglement after collapse is nonzero only at times where the partial derivative curve is positive (since $\textit{a}$ is positive, the relation in (\ref{eita-asol-byapar}) holds). \(\gamma = \frac{1}{2}\), as before.}
\label{fig:4}
\end{figure}

Already in Fig.\,\ref{fig:2}, we see that QD behaves differently for times for which entanglement revival takes place. To get a clearer picture of the situation, in Fig.\,\ref{fig:3}, we plot entanglement and QD for different fixed times $\tilde{t}$, as functions of the  
initial field strength $\tilde{a}$. Only for those times \(\tilde{t}\) for which QD is an increasing function of the field strength \(\tilde{a}\) at \(\tilde{a} = \tilde{a}_c\), \(\tilde{a}_c\) being the field strength at which entanglement vanishes for a given \(\tilde{t}\), there is a revival of entanglement at a higher value of \(\tilde{a}\), as stated in (\ref{eita-asol-byapar}).
In Fig.\,\ref{fig:4}, we plot the partial derivative of QD with respect to \(\tilde{a}\), at \(\tilde{a} = \tilde{a}_c\). It is a 
pictorial representation of the relation given in (\ref{eita-asol-byapar}). A similar conjecture can also be obtained by studying the correlation dynamics by varying the anisotropy parameter \(\gamma\) even though no consistent DPT is observed for such evolutions.

\subsection{Behavior of Quantum Work-Deficit}
\begin{figure}[htb]
\begin{center}
\epsfig{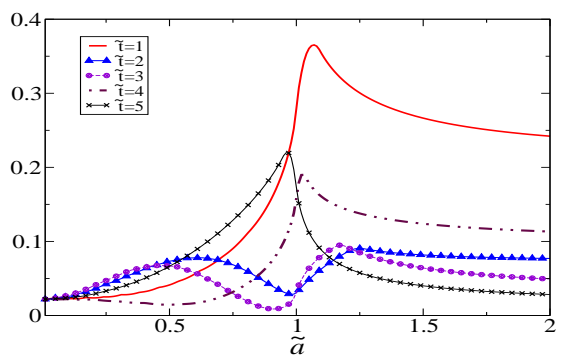}
\end{center}
\caption{(Color online) Quantum work-deficit (measured in qubits) is plotted against the initial field strength \(\tilde{a}\) (dimensionless) for different times \(\tilde{t}\) (dimensionless). Around times \(\tilde{t}=1\) and \(\tilde{t}=4\), for which entanglement reappearance is observed, QWD rises to high values for moderately large \(\tilde{a}\). Again, \(\gamma = \frac{1}{2}\).}
\label{fig:6}
\end{figure}

We consider a second information-theoretic measure of quantum correlations, QWD. Sections of its plot at different times is given in Fig.\,\ref{fig:6}, and it is seen that around times \(\tilde{t}=1\) and \(\tilde{t}=4\), QWD reaches high values for moderately large
\(\tilde{a}\). Hence even though no working relation, unlike for QD, exists for QWD, it carries a predictive capacity, which will be explored in a future work \cite{inprep}.

\section{Discussions}
We have shown that a measure of quantum correlations known as QD, that goes beyond the standard entanglement-separability paradigm, 
can be used to encapsulate a criterion that tells us when entanglement will revive after collapse in an infinite anisotropic quantum XY spin chain. Indeed, we have shown that an increase of QD around entanglement collapse indicates a revival of entanglement at higher fields. This result could potentially be useful in applications of entanglement as a resource in quantum communication and other quantum information tasks to indicate which parameter zones of a given physical system contain entangled states, and which do not. 

We would like to test whether the findings are generic, in the sense that other measures of quantum correlations could show the same behavior, by considering a 
second information-theoretic measure of quantum correlations, QWD. We have seen that even though no working relation, unlike for QD, exists for QWD, it carries a 
predictive capacity which could be a generic feature of other measures of quantum correlations defined beyond the entanglement-separability criteria.

Studies in information-theoretic measures of quantum correlations, such as QD and QWD, have revealed that such measures give a fine-grained picture of quantum  states of distributed systems in comparison to that provided by entanglement. Our study indicates that such a fine-grained picture can show the underlying reason for the dynamics seen for entanglement in quantum many-body systems. 

\acknowledgments
The work of HSD is supported by the University Grants Commission (UGC), India. 
HSD also thanks the Harish-Chandra Research Institute (HRI) for hospitality and support during visits. 
We acknowledge computations performed at the cluster computing facility in the HRI (http://cluster.hri.res.in/), and also at the 
UGC-DRS computing facility at the Jawaharlal Nehru University (JNU). ASD and US thank the JNU for hospitality.

\end{document}